\documentclass{elsart}
\usepackage{makeidx}
\usepackage{amssymb}
\usepackage{graphicx}

\begin{document}

\begin{frontmatter}

\title{\textbf{Incidence of nonextensive thermodynamics in temporal scaling
at Feigenbaum points}}
\author{A. Robledo}
\ead{robledo@fisica.unam.mx }
\address{Instituto de F\'{\i}sica,\\
Universidad Nacional Aut\'{o}noma de M\'{e}xico,\\
Apartado Postal 20-364, M\'{e}xico 01000 D.F., Mexico.}

\begin{abstract}
Recently, in Phys. Rev. Lett. 95, 140601 (2005), P. Grassberger
addresses the interesting issue of the applicability of
$q$-statistics to the renowned Feigenbaum attractor. He concludes
there is no genuine connection between the dynamics at the
critical attractor and the generalized statistics and argues
against its usefulness and correctness. Yet, several points are
not in line with our current knowledge, nor are his
interpretations. We refer here only to the dynamics \textit{on}
the attractor to point out that a correct reading of recent
developments invalidates his basic claim.
\end{abstract}

\end{frontmatter}


\section{Introduction}

A sharp disagreement and refutation has been published recently \cite%
{grassberger1} on the contention, based on many studies \cite{tsallis1} -
\cite{robledo7}, that the generalized statistical-mechanical scheme known as
nonextensive statistical mechanics \cite{tsallis9} is pertinent to the
dynamical properties at the so-called onset of chaos in dissipative
low-dimensional iterated maps. The `dialogue' between advocates and
opponents of the appropriateness of this new formalism, here referred to as $%
q$-statistics, to this old problem has been hampered by the lack of both: i)
an all-inclusive explanation of how and why the new formalism relates to the
aforementioned dynamics, and ii) an objective appraisal of the published
results on the subject as they have developed in time and in depth. Whereas
a definite rationalization may be hastened by the ongoing accumulation of
information from current research, unbalanced judgments are sustained by the
prevalent belief that, after the historic and extensive work on the
transitions to chaos carried out a few decades ago, there are no major
unknown features left to be revealed, nor intrinsic value in new approaches.
Here we attempt to break this impasse by explaining as plainly as possible
the manner in which $q$-statistics manifests at the transitions to chaos,
and by indicating how it relates to previous approaches, such as the
`thermodynamic formalism' for nonlinear dynamics as adapted by Mori and
colleagues \cite{mori1} to the study of this kind of attractor.

When the control parameter of a nonlinear one-dimensional map is set at the
threshold between periodic and chaotic motion\ an unusual and complicated
dynamics arises, whose features have been explored long ago \cite%
{grassberger2}-\cite{mori3} and described recently in greater detail \cite%
{robledo0} - \cite{robledo7}. For both the period doubling and the
quasiperiodic routes to chaos the transition from periodic to chaotic
behavior is mediated by the appearance of a multifractal `critical'
attractor with geometrical properties already known for a couple of decades
\cite{schuster1}-\cite{hilborn1}. Critical attractors have a vanishing
Lyapunov coefficient $\lambda _{1}$ and a sensitivity to initial conditions $%
\xi _{t}$ that does not converge to any single-valued function but instead
displays a fluctuating pattern that grows as a power law in time $t$ \cite%
{mori1}. Trajectories \textit{within} such a critical attractor show
self-similar temporal structures, they preserve memory of their previous
locations and do not have the mixing property of truly chaotic trajectories
\cite{mori1}. A special version of the thermodynamic formalism for
deterministic chaos \cite{beck1}, \cite{halsey1}-\cite{paladin1} was adapted
long ago \cite{mori1} to study the dynamics at critical attractors and
quantitative results were obtained that provided a first understanding,
particularly about the envelope of the fluctuating $\xi _{t}$ and the
occurrence of a so-called `$q$-phase' dynamical phase transition \cite{mori1}%
.

Below we describe the scaling properties of the fluctuating sensitivity $\xi
_{t}(x_{in})$ \textit{at} a multifractal critical attractor as well as those
of its associated spectra of \textit{generalized} Lyapunov coefficients $%
\lambda (x_{in})$ (where $x_{in}$ stands for the initial position of
trajectories). We illustrate these properties and their connection to the $q$%
-statistical expressions by considering mainly the specific case of the
period doubling onset of chaos, i.e. the so-called Feigenbaum attractor. We
explain that the dynamics at the attractor consists of families of dynamical
$q$-phase transitions and discuss the ensuing relationship between the
thermodynamic and the $q$-statistical formalisms.

We refer to various points raised in Ref. \cite{grassberger1} as a means of
clarifying our current understanding of this problem \cite{robledo8}. We
chose this format in part because subsequent critical opinions \cite%
{nauenberg1} on our developments \cite{robledo0} - \cite{robledo7} appear to
be based on a cursory and unquestioning belief that the arguments in Ref.
\cite{grassberger1} are indisputable and its views definitive. However, as
we demonstrate below, the contents in Ref. \cite{grassberger1} suffer
shortcomings that seem to stem from overconfidence and poor reading of the
studies cited there, particularly \cite{robledo0} - \cite{robledo7}. Another
reason for adopting an extended `comment' layout is that the customary
channels for response have been discouraged or barred by the editors of the
journal and magazine where the criticisms have been published, preventing an
informed, in-depth and balanced debate on this issue. We would wish that
this propagation of swiftly-taken and disingenuous judgements be
counterbalanced by standard objective reasoning.

In Section 2 we sum up the features of $q$-statistics in relation to the
dynamics we discuss. In Section 3 we explain the structure of $\xi _{t}$ for
the fluctuating dynamics at the Feigenbaum attractor. We point out there
that this dynamical structure is distinguished by a built-in `aging' or
`waiting time' scaling law. In Section 4\ we describe the relationship
between the thermodynamic approach and $q$-statistics as applied to critical
attractors. We indicate that the fixed values for the Tsallis' $q$ index
correspond to the values that Mori's field variable $\mathsf{q}$ takes at
the dynamical phase transitions, i.e. $\mathsf{q}_{trans}=q$. In Section 5
we demonstrate the condition for the linear growth of the $q$-entropy $S_{q}$%
, i.e. the occurrence of $q$-phase transitions. We also prove the identity
between the rates of $q$-entropy growth and the $q$-generalized Lyapunov
coefficients. For clarity of presentation\ the previous sections focus on
the properties that link only the two dominant scaling regions of the
multifractal attractor, its most crowded and most sparse. In Section 6 all
the scaling regions of the multifractal are considered, it is shown how $\xi
_{t}$ is obtained in terms of the discontinuities of Feigenbaum's trajectory
scaling function $\sigma $, and it is observed that the entire dynamics is
made of a hierarchical family of pairs of $q$-phase transitions. In Section
7 we refer to the parallelisms in the expression of $q$-statistics in the
other two routes to chaos displayed by low-dimensional maps, via
quasiperiodicity and via intermittency. In Section 8 we summarize the main
results.

\section{$q$-statistics for critical attractors}

First of all it is indispensable to stipulate what (according to the author)
is intended by the manifestation of $q$-statistics in the dynamics at a
critical attractor. This can be summarized in two related properties. The
first concerns the (finite time) sensitivity to initial conditions $\xi _{t}$%
, defined as
\begin{equation}
\xi _{t}(x_{in})\equiv \lim_{\Delta x_{0}\rightarrow 0}\left\vert \frac{%
\Delta x_{t}}{\Delta x_{in}}\right\vert ,  \label{sensitivity}
\end{equation}%
where $\Delta x_{in}$ is the initial separation of two trajectories and $%
\Delta x_{t}$ that at time $t$. This quantity is expected to involve
expressions of the form%
\begin{equation}
\xi _{t}(x_{in})=\exp _{q}[\lambda _{q}(x_{in})\ t],  \label{sensitivity1}
\end{equation}%
where $q$ is the entropic index, $\lambda _{q}(x_{in})$ is the $q$%
-generalized Lyapunov coefficient, and $\exp _{q}(x)\equiv \lbrack
1-(q-1)x]^{-1/(q-1)}$ is the $q$-exponential function. The second property
relates to `temporal extensivity' of entropy production \cite{abe1} at
critical attractors \cite{tsallis1}, \cite{robledo4}, \cite{robledo6}, \cite%
{robledo7}. That is, linear growth with time $t$ of the entropy associated
to an ensemble of trajectories. The fitting expression for the rate of
entropy production $K_{q}(x_{in})$ to be used is thought to be given by
\begin{equation}
K_{q}(x_{in})\ t=S_{q}(t,x_{in})-S_{q}(0,x_{in}),  \label{q-entropyrate1}
\end{equation}%
where
\begin{equation}
S_{q}\equiv \sum_{i}p_{i}\ln _{q}\ p_{i}^{-1}=\frac{1-\sum_{i}^{W}p_{i}^{q}}{%
q-1},  \label{tsallisentropy1}
\end{equation}%
is the Tsallis entropy \cite{tsallis9} and where $p_{i}(t)$ is the
distribution of the trajectories in the ensemble at time $t$ given that they
were distributed initially within a small interval around $x_{in}$. Recall
that $\ln _{q}y\equiv (y^{1-q}-1)/(1-q)$ is the inverse of $\exp _{q}(y)$
and notice the explicit dependence of all quantities on the initial position
$x_{in}$.

Several comments are in order:

i) It is anticipated that the index $q$ in Eq. (\ref{sensitivity1}) takes a
well defined value determined by the basic attractor properties such as its
universal constants. For a multifractal critical attractor there may be a
distinct discrete family of such values for $q$.

ii) Because of the memory retention of the trajectories the dependence on $%
x_{in}$ in Eq. (\ref{sensitivity1}) (or Eq. (\ref{q-entropyrate1})) does not
disappear for sufficiently large $t$. Notice that the usual large $t$
limiting condition is not present in the definition of $\xi _{t}$ in Eq. (%
\ref{sensitivity1}) nor in Eq. (\ref{q-entropyrate1}).

iii) Since\ $\xi _{t}$ fluctuates according to a deterministic pattern
(given by the specific route to chaos studied) the variable $t$ in Eq. (\ref%
{sensitivity1}) cannot run through all positive integers sequentially but
only through specific infinite subsets. However, and as seen below, all
times $t$ appear in Eq. (\ref{sensitivity1}) when $x_{in}$ is varied.

iv) The standard exponential form for $\xi _{t}$ with the ordinary $\lambda
_{1}$ is obtained from Eq. (\ref{sensitivity1}) when $q\rightarrow 1$. This
limit is brought about when the control parameter in the map is shifted from
its value at the chaos threshold to a value that corresponds to either a
periodic or a chaotic attractor. For these cases the fluctuations in $\xi
_{t}$ die out and the Lyapunov spectra $\lambda _{q}(x_{in})$ collapses into
the single number $\lambda _{1}$ independent of $x_{in}$ for $t\;$large.

v) Depending on the value of $q$ and the sign of $\lambda _{q}(x_{in})$, the
$q$-exponential sensitivity can grow or decrease asymptotically as a power
law with $t$, but it can also grow or decrease faster than an ordinary
exponential. As we see below, for a multifractal critical attractor Eq. (\ref%
{sensitivity1}) is obtained from an exact transformation of a pure power law
\cite{robledo3}, \cite{robledo4}, but a faster than exponential $q$%
-sensitivity is rigorously obtained for the critical attractor at a tangent
bifurcation \cite{robledo2}.

vi) The value of the index $q$ that makes the rate $K_{q}$ a time
independent constant is expected to be the same value appearing in Eq. (\ref%
{sensitivity1}).

vii) The identity $K_{q}(x_{in})=\lambda _{q}(x_{in})$ is supposed to hold.
In the limit $q\rightarrow 1$ the familiar identity $K_{1}=$ $\lambda _{1}$
\cite{hilborn1} is recovered (where the rate of entropy production $K_{1}$
is given by $K_{1}t=S_{1}(t)-S_{1}(0)$ and $S_{1}=-\sum_{i}p_{i}\ln p_{i}$).
Note that in this identity the rate $K_{1}$ is not the so-called KS entropy
\cite{schuster1}-\cite{hilborn1} but a closely related quantity. See Section
5.

\section{Fluctuating dynamics at the Feigenbaum attractor}

Let us consider now properties specific to the Feigenbaum map $g(x)$,
obtained from the fixed point equation $g(x)=\alpha g(g(x/\alpha ))$ with $%
g(0)=1$ and $g^{\prime }(0)=0$, and where $\alpha =-2.5029079...$ is one of
Feigenbaum's universal constants \cite{schuster1}. For expediency we shall
from now on denote the absolute value $\left\vert \alpha \right\vert $ by $%
\alpha $. Numerically, the properties of $g(x)$ can be conveniently obtained
from the logistic map
\begin{equation}
f_{\mu ,2}(x)=1-\mu x^{2},\;-1\leq x\leq 1,  \label{logistic1}
\end{equation}%
with $\mu =$ $\mu _{\infty }=1.401155189...$. A first important issue that
merits amplification and explanation lies behind Grassberger's comment that
his Eq. (3) in \cite{grassberger1}, here rewritten as%
\begin{equation}
\xi _{t}(x_{in})=\alpha ^{k}  \label{sensitivity2}
\end{equation}%
holds only for special values of time, e.g. $t=2^{k}-1$, $k=0,1,...$ when $%
x_{in}=1$. Not quite, there are many other time sequences all of which
satisfy Eq. (\ref{sensitivity2}) \textit{exactly }\cite{robledo4}. To see
this consider the trajectory $x_{n}$ with initial position $x_{in}=0$,
plotted in Fig. 1 in a way that makes evident the structure of the time
sequences and the power law property that their positions share. It can be
observed there that the positions $x_{n(k,l)}$ of the sequences with times $%
n(k,l)=(2l+1)2^{k}$, each obtained by running through $k=0,1,2,...$ for a
fixed value of $l=0,1,2,...$ fall along straight diagonal lines. The
starting elements of these sequences are the positions $x_{2l+1}$. To be
precise, the \textit{entire} attractor can be decomposed into this family of
position sequences as every natural number $n$ appears in one of them.
Actually, the complete fluctuating sensitivity $\xi _{t}(x_{in}=1)$ can be
decomposed into a family of identical power laws given by Eq. (\ref%
{sensitivity2}) with $t=n-(2l+1)$ \cite{robledo4}. The sequence for $l=0$
belongs to the envelope of $\xi _{t}$ while those for other values of $l$
are shifted replicas that describe interior values of $\xi _{t}$. This
feature offers an enormous simplification to the study of the dynamics on
the attractor. Time rescaling of the form $t/t_{w}$ with $t_{w}=2l+1$ makes
all power law sequences collapse into a single one in a manner analogous to
the so-called `aging' property of `glassy' dynamics where $t_{w}$ is a
`waiting time' \cite{robledo9}, \cite{robledo10}. \newline
\begin{figure}[tbph]
\centering\includegraphics[width=9cm]{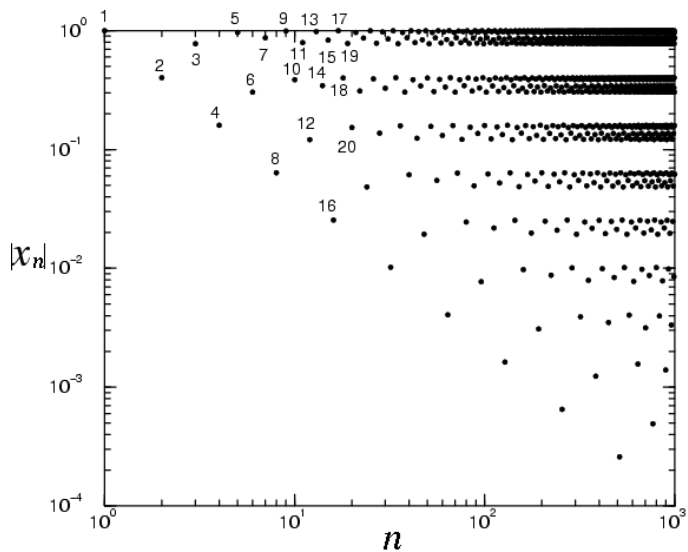} 
\caption{{\protect\small Absolute values of positions in logarithmic scales
of the first 1000 iterations for the trajectory of the logistic map at the
onset of chaos }$\protect\mu _{\infty }${\protect\small \ with initial
condition }$x_{in}=0${\protect\small . The numbers correspond to iteration
times. The power-law decay shared by the time sequences mentioned in the
text can be clearly appreciated.} }
\end{figure}

The family of power laws in Eq. (\ref{sensitivity2}) can be rewritten \cite%
{robledo4}, with the use of the identity%
\begin{equation}
\alpha ^{k}\equiv \left( 1+\frac{t}{2l+1}\right) ^{\ln \alpha /\ln 2},
\label{identity1}
\end{equation}%
with $t=$ $(2l+1)2^{k}-2l-1$, as the family of $q$-exponentials,%
\begin{equation}
\xi _{t}(x_{in})=\exp _{q}[\lambda _{q}^{(l)}t],  \label{sensitivity3}
\end{equation}%
\textit{all} with the same value of $q$. Above, $q=1-\ln 2/\ln \alpha $ and $%
\lambda _{q}^{(l)}=(2l+1)^{-1}\ln \alpha /\ln 2$ with $t=$ $n-2l-1$, when $%
x_{in}=1$, or with $t=n=$ $(2l+1)2^{k}$ when $x_{in}$ is any of the
positions $x_{2l+1}$ in Fig. 1, in all cases running through $k=0,1,...$
while $l$ is fixed to a given value $l=0,1,...$ Alternatively, Eq. (\ref%
{sensitivity3}) can be written as the single $q$-exponential%
\begin{equation}
\xi _{t}(x_{in})=\exp _{q}[\lambda _{q}^{(0)}\frac{t}{t_{w}}].
\label{sensitivity4}
\end{equation}%
So, contrary to the statement in \cite{grassberger1} - that his Eq. (3) is
\textit{not} a scaling law because it only holds for special values of $t$ -
Eq. (\ref{sensitivity2}), and its equivalent forms Eqs. (\ref{sensitivity3})
or (\ref{sensitivity4}), in fact correspond to a more complex form of a
scaling law, one that fits the fluctuating dynamics at the attractor, and
referred here as a 'two-time' scaling law. Below we make clear that the
positions of these sequences for large $k$ belong to one scaling region
within the multifractal, its most sparse region.

A persuasive reason for considering a $q$-exponential as a further way to
write a power law is the presence of a time scale factor: the generalized
Lyapunov coefficient $\lambda _{q}^{(l)}$ that appears multiplying $t$ in $%
\xi _{t}$. This useful quantity (hidden in the power law) can be immediately
`read' from the anomalous $\xi _{t}$ just as the ordinary Lyapunov
coefficient $\lambda _{1}$ is read from the exponential $\xi _{t}$ of
chaotic dynamics. It is important to stress that it \textit{is} in the
scaling limit (large $n$ or $k$) that $\xi _{t}$ becomes $\alpha ^{k}$, and
it is \textit{this} which transforms (exactly) into the $q$-exponential $\xi
_{t}$. Evidently, reference and use of this property involves more than a
mere choice between a power law and a $q$-exponential representation of the
sensitivity. The beauty of a scale-free power law is there to enjoy. As for
generality, all of these properties apply with only minor changes to the
period-doubling transition to chaos in unimodal maps with extremum of order $%
z>1$ \cite{robledo6}. The derivation of Eq. (\ref{sensitivity2}) for $z>1$
is given in \cite{robledo5} so there is no need to reproduce it in \cite%
{grassberger1}.

So, what is the advantage in studying the nonmixing trajectories at the
Feigenbaum attractor in terms of the above-explained power-law network
structure for $\xi _{t}$ instead of the time and position averages described
in \cite{grassberger1}? Clearly, the dynamical organization within the
attractor is hard to resolve from a simple time evolution: starting from an
arbitrary position $x_{in}$ \textit{on} the attractor and recorded at every $%
t$. What is observed are strong fluctuations with a scrambled structure that
persist in time. Conversely, unsystematic averages over $x_{in}$ and/or $t$
would rub out details of the multiscale properties. However, if specific
initial positions with known location within the multifractal are chosen,
and subsequent positions are observed \textit{only} at pre-selected times,
e.g. $t=2^{k}-1$, when the trajectories visit another region of choice, a
well-defined $q$-exponential sensitivity $\xi _{t}$ appears, with $q$ and
the Lyapunov spectrum $\lambda (x_{in})$ fixed by the attractor universal
constants.

\section{Thermodynamic approach and $q$-statistics}

We now address the connection between the thermodynamic approach as employed
by Mori and colleagues \cite{mori1} to study dynamical phase transitions at
critical attractors and the $q$-statistical properties of the Feigenbaum
attractor. In particular: i) We establish the equivalence of the generalized
Lyapunov coefficients present in the two methods. ii) We show the link
between the occurrence of dynamical phase transitions and the existence of
unique values for the entropic index $q$. iii) We demonstrate the linear
growth of the Tsallis entropy at the dynamical phase transitions and the
identity between the entropy growth rates with their corresponding $q$%
-Lyapunov coefficients.

\subsection{ Generalized Lyapunov coefficients and $q$-deformation}

We make use of the scaling features of $\xi _{t}$ just described to examine
Grassberger's advice in \cite{grassberger1} that Eq. (4) in \cite%
{grassberger1} gives the natural generalization of the Lyapunov coefficient $%
\lambda _{1}$. Indeed, a closely related prescription,
\begin{equation}
\lambda _{t}(x_{in})\equiv \frac{1}{\ln t}\ln \left\vert \frac{%
dg^{(t)}(x_{in})}{dx_{in}}\right\vert ,  \label{lyapunov1}
\end{equation}%
was initially adopted in Refs. \cite{politi1}, \cite{mori3}. It differs
however from Eq. (4) in \cite{grassberger1} in that no time average is
taken. As a function of consecutive values of $t$, $\lambda _{t}(x_{in})$
does not converge to any constant but does undergo recurrence. This
recurrence suggests inspection of $\lambda _{t}(x_{in})$ for values of $t$
along the time sequences described in the previous section. A
straightforward calculation shows that $\lambda _{t}(x_{in}=1)$ as defined
in Eq. (\ref{lyapunov1}) is exactly given by%
\begin{equation}
\lambda _{t}(x_{in}=1)=\frac{1}{t}\ln _{q}\left\vert \frac{dg^{(t)}(x_{in})}{%
dx_{in}}\right\vert _{x_{in}=1}=\frac{\ln \alpha }{(2l+1)\ln 2}=\lambda
_{q}^{(l)},  \label{lyapunov2}
\end{equation}%
where $t$, as before, runs through the sequences $t=$ $(2l+1)2^{k}-2l-1$, $%
k=0,1,...$, with $l=0,1,...$ fixed. So, the earlier definition for the
generalized Lyapunov coefficient \textit{is equivalent} to that given for
the same quantity by the $q$-statistics. The meaning of the index $q$ is
given above. It is the degree of `$q$-deformation' of the ordinary logarithm
that makes $\lambda _{t}$ finite for large $t$. It is not difficult to
corroborate that this result is valid for general $x_{in}$ on the attractor.

\subsection{Dynamical partition function}

Eqs. (\ref{lyapunov1}) and (\ref{lyapunov2}) suggest a broader connection
between the thermodynamic formalism and the $q$-statistical features found
for the dynamics at the Feigenbaum attractor. Indeed, it has been shown in
\cite{robledo6}, \cite{robledo7} that the physical origin of the existence
of a distinct value (or values) for the entropic index $q$ is linked to the
occurrence of dynamical phase transitions, of Mori's $q$-phase type \cite%
{mori1}, which connect qualitatively different regions of the attractor.

The thermodynamic formalism is built on the dynamical partition function%
\begin{equation}
Z(n,\mathsf{q})\equiv \int d\lambda \ P(\lambda ,t)\ W(\lambda ,t)^{1-%
\mathsf{q}},  \label{partition1}
\end{equation}%
where the weight $W(\lambda ,t)$ was assumed \cite{mori1} to have the form $%
W(\lambda ,t)=\exp (\lambda t)$ for chaotic attractors and $W(\lambda
,t)=t^{\lambda }$ for the attractor at the onset of chaos, with $\lambda $ a
generalized Lyapunov coefficient (whose dependence on $t$ and $x_{in}$ is
not written explicitly). The variable $\mathsf{q}$ plays the role of a
`thermodynamic field' like the external magnetic field in a thermal magnet
or, alternatively, $1-\mathsf{q}$ can be thought analogous to the inverse
temperature. We note that $W(\lambda ,t)$ can be written in both cases as%
\begin{equation}
W(\lambda ,t)=\left\vert \frac{df^{(t)}(x_{in})}{dx_{in}}\right\vert ,
\label{weight1}
\end{equation}%
where $f(x)$ is the iterated map under consideration, that is,
\begin{equation}
\lambda _{t}(x_{in})=B(t)\ln \left\vert \frac{df^{(t)}(x_{in})}{dx_{in}}%
\right\vert  \label{lyapunov2a}
\end{equation}%
with $B(t)=$ $t^{-1}$ for chaotic attractors and $B(t)=(\ln t)^{-1}$ for the
fluctuating sensitivity at the critical attractor \cite{mori1}. As
mentioned, for chaotic attractors the fluctuations of $\lambda _{t}(x_{in})$
die out as $t\rightarrow \infty $ and this quantity becomes the ordinary
Lyapunov coefficient $\lambda _{1}$ independent of the initial position $%
x_{in}$. For critical attractors $\lambda _{t}(x_{in})$ maintains its
dependence on $t$ and $x_{in}$ but as already seen it is a well-defined
constant \textit{provided} $t$ takes values along each time sequence, or if
the two-time scaling property (Eq. (\ref{sensitivity4})) is \textit{summoned}%
.

For critical attractors the density distribution for the values of $\lambda $%
, $t\gg 1$, $P(\lambda ,t)$, is written in the form \cite{mori1}
\begin{equation}
P(\lambda ,t)=t^{-\psi (\lambda )}P(0,t),  \label{distribution1}
\end{equation}%
where $\psi (\lambda )$ is a concave spectrum of the fluctuations of $%
\lambda $ with minimum $\psi (0)=0$ and is obtained as the Legendre
transform of the `free energy' function $\phi (\mathsf{q})$, defined as $%
\phi (\mathsf{q})\equiv -\lim_{t\rightarrow \infty }\ln Z(t,\mathsf{q})/\ln
t $. The generalized Lyapunov coefficient $\lambda (\mathsf{q})$ is given by
$\lambda (\mathsf{q})\equiv d\phi (\mathsf{q})/d\mathsf{q}$ \cite{mori1}.
The functions $\phi (\mathsf{q})$ and $\psi (\lambda )$ are the dynamic
counterparts of the Renyi dimensions $D(\mathsf{q})$ and the spectrum $f(%
\widetilde{\alpha })$ that characterize the geometric structure of the
attractor \cite{beck1}, \cite{hilborn1}. For non hyperbolic attractors like
those at the onset of chaos the functions $\phi (\mathsf{q})$ and $\psi
(\lambda )$ that are used to describe the dynamics are independent of $D(%
\mathsf{q})$ and $f(\widetilde{\alpha })$.

\subsection{$q$-phase transitions and Tsallis $q$ index}

As with thermal 1st order phase transitions, a $q$-phase\ transition is
indicated by a section of linear slope $m_{c}=1-q$ in the spectrum (free
energy) $\psi (\lambda )$ and consequently a discontinuity at $\mathsf{q}=q$
in the Lyapunov function (order parameter) $\lambda (\mathsf{q})$. For the
Feigenbaum attractor a single $q$-phase transition was numerically
determined \cite{mori3} and found to occur approximately at a value around $%
m_{c}=-(1-q)\simeq -0.7$. It was pointed out in Ref. \cite{politi1}, \cite%
{mori3} that this value would actually be $m_{c}=-(1-q)=-\ln 2/\ln \alpha
=-0.7555...$

From the knowledge we have gained on $q$-generalized Lyapunov coefficients
for this attractor, e.g. Eq. (\ref{lyapunov2}), we can determine the free
energy $\psi (\lambda )$. But to do this we need to extend our results a bit
further. We note that $\lambda _{t}(x_{in})$ in Eq. (\ref{lyapunov2}),
corresponds to trajectories that originate in the most crowded region of the
attractor, e.g. $x_{in}=1$, and terminate at its most sparse region, $%
x_{n=(2l+1)2^{k}}\simeq 0$, large $k$. Naturally, these trajectories when
observed at times $n=(2l+1)2^{k}$ grow apart from each other as $k$
increases ($\lambda _{t}(x_{in})>0$). When the inverse situation is
considered, e.g. $x_{in}=0$, with final positions $x_{n=(2l+1)2^{k}+1}\simeq
1$, one obtains \cite{robledo6}%
\begin{equation}
\xi _{t}(x_{in})=\exp _{Q}[\lambda _{Q}^{(l)}t],  \label{sensitivity5}
\end{equation}%
with $Q=2-q=1+\ln 2/\ln \alpha $ and%
\begin{equation}
\lambda _{t}(x_{in}=0)=-\frac{2\ln \alpha }{(2l+1)\ln 2}=\lambda _{Q}^{(l)},
\label{lyapunov3a}
\end{equation}%
where $t=$ $(2l+1)2^{k}-2l$, $k=0,1,...$, $l=0,1,...$ Above, of course, $%
\lambda _{t}(0)<0$ since the trajectories when observed at times $%
(2l+1)2^{k}+1$ come progressively close to each other. Notice that the
change in sign of $\lambda _{t}$ and the relation $Q=2-q$ match the property
of inversion of the $q$-exponential, $\exp _{q}(x)=1/\exp _{2-q}(-x)$. Also,
trivially, when the initial and final positions of trajectories belong to
the same region within the attractor, e.g. $x_{1}=1$ and $%
x_{n=(2l+1)2^{k}+1}\simeq 1$, one has $\xi _{t}=1$ with $\lambda _{q}=0$.

From the results for $\lambda _{q}^{(l)}$ and $\lambda _{Q}^{(l)}$ the
Lyapunov function $\lambda (\mathsf{q})$ and spectrum $\psi (\lambda )$ are
constructed \cite{robledo6}
\begin{equation}
\lambda (\mathsf{q})=\left\{
\begin{array}{l}
\lambda _{q}^{(0)},\;-\infty <\mathsf{q}\leq q, \\
0,\;\;\;\;\;q<\mathsf{q}<Q, \\
\lambda _{Q}^{(0)},\;\;\;Q\leq \mathsf{q}<\infty ,%
\end{array}%
\right.  \label{lyapunov4}
\end{equation}%
and
\begin{equation}
\psi (\lambda )=\left\{
\begin{array}{l}
(1-Q)\lambda ,\;\lambda _{Q}^{(0)}<\lambda <0, \\
(1-q)\lambda ,\;0<\lambda <\lambda _{q}^{(0)}.%
\end{array}%
\right. ,  \label{spectrum1}
\end{equation}%
with $\lambda _{q}^{(0)}=\ln \alpha /\ln 2\simeq 1.323$ and $\lambda
_{Q}^{(0)}=-2\lambda _{q}^{(0)}\simeq -2.646$. The constant slopes of $\psi
(\lambda )$ represent the $q$-phase transitions associated to trajectories
linking two regions of the attractor, $x\simeq 1$ and $x\simeq 0$, and their
values $1-q$ and $q-1$ correspond to the Tsallis index $q$ obtained from $%
\xi _{t}$. The slope $q-1\simeq -0.7555$ coincides with that initially
detected in Refs. \cite{mori2}, \cite{politi1}.

The significance of the above exercise is that it reveals the physical
reason for the existence of a well-defined value of the entropic index $q$
in critical attractor dynamics. This is the occurrence of a $q$-phase
transition. The value of the field $\mathsf{q}$ at which the transition
takes place is precisely $q$ (and the same of course for the inverse
process, $\mathsf{q}_{trans}=Q$).

\section{Temporal extensivity of the $q$-entropy}

As we have seen, along each time sequence $n(k,l)=$ $(2l+1)2^{k}$, $l$
fixed, the sensitivity $\xi _{t}$ with $x_{in}=1$ is given by the $q$%
-exponential in Eq. (\ref{sensitivity3}), and consequently we can express $%
\lambda _{t}(x_{in})$ in terms of the $q$-deformed logarithm in Eq. (\ref%
{lyapunov2}). Therefore for times of the form $t=n-(2l+1)$ we have%
\begin{equation}
P(\lambda ,t)W(\lambda ,t)=\delta (\lambda -\lambda _{q}^{(l)})\exp
_{q}(\lambda _{q}^{(l)}t),  \label{weight2}
\end{equation}%
and
\begin{equation}
Z(t,\mathsf{q})=W(\lambda _{q}^{(l)},t)^{1-\mathsf{q}}=\left[ 1+(1-q)\lambda
_{q}^{(l)}t\right] ^{(1-\mathsf{q})/(1-q)}.  \label{partition2}
\end{equation}%
Notice that in Eq. (\ref{partition2}) $\mathsf{q}$ is a running variable
while $q=1-\ln 2/\ln \alpha $ is fixed.

Our next step is to consider the uniform probability distribution, for fixed
$\lambda _{q}^{(l)}$ and $t$, given by $p_{i}(\lambda _{q}^{(l)},t)=%
\overline{W}^{-1}$, $i=1,...,\overline{W}$, where $\overline{W}$ is the
integer nearest to (a large) $W$. A trajectory starting at $x_{in}=1$ visits
a new site $x_{n}$ belonging to the time sequence $n(k,l)$, $l$ fixed, every
time the variable $k$ increases by one unit and never repeats one. Each time
this event occurs phase space is progressively covered with an interval of
length $\Delta x_{k,l}=x_{(2l+1)2^{k}}-x_{(2l+1)2^{k+1}}>0$. The difference
of the logarithms of the times between any two such consecutive events is
the constant $\ln 2$, $n$ large, whereas the logarithm of the corresponding
phase space distance covered $\Delta x_{k,l}$ is the constant $\ln \alpha $.
We define $W_{k}$ to be the total phase space distance covered by these
events up to time $n=$ $(2l+1)2^{k}$. In logarithmic scales this is a linear
growth process in space and time from which the constant probability $%
\overline{W}_{k}^{-1}$ of the uniform distribution $p_{i}$ is defined. In
ordinary phase space $x$ and time $t$ we obtain instead $W_{t}^{-1}=\exp
(-\lambda _{q}^{(l)}t)$. We then have that
\begin{equation}
Z(t,\mathsf{q})=\sum_{i=1}^{\overline{W}}[p_{i}(t)]^{\mathsf{q}}=1+(1-%
\mathsf{q})S_{\mathsf{q}}(t),  \label{partition3}
\end{equation}%
where $S_{\mathsf{q}}=\ln _{\mathsf{q}}\overline{W}$ is the Tsallis entropy
for $p_{i}$.

The usefulness of the $q$-statistical approach is now evident when we recall
from our discussion above that the dynamics \textit{on} the critical
attractor displays as its main feature a $q$-phase transition and that
\textit{at} this transition the field variable $\mathsf{q}$ takes the
specific value $\mathsf{q}_{trans}=q$. Comparison of Eqs. (\ref{partition2})
and (\ref{partition3}) indicates that at each such transition both $Z$ and $%
S_{q}$ grow \textit{linearly} with time along the sequences $n(k,l)$. In
addition, $\lambda _{q}^{(l)}$ can be determined from $S_{q}/t=\lambda
_{q}^{(l)}$. Thus, with the knowledge we have gained, a convenient procedure
for determining the relevant quantities for the fluctuating dynamics at a
critical multifractal attractor could take advantage of the above
properties. Curiously, the same features have been observed for $S_{q}$ with
$q=0$ associated to trajectories (with linear $\xi _{t}$) in a conservative
two-dimensional map with vanishing ordinary Lyapunov coefficients \cite%
{tsallis10}.

Further, the identity $S_{q}/t=\lambda _{q}^{(l)}$ (also derived in \cite%
{robledo6}) between the rate of $q$-entropy change and the generalized $%
\lambda $ is not the identity
\begin{equation}
t^{-1}(S_{1}(t)-S_{1}(0))=t^{-1}\ln \left\vert \frac{dg^{(t)}(x_{in})}{%
dx_{in}}\right\vert  \label{bgrate1}
\end{equation}%
in \cite{grassberger1} (the zero identity for $t\rightarrow \infty $) but
refers to $\lambda _{t}(x_{in})$ as above. Yes, it considers an
instantaneous entropy rate but it is comparable in the sense of \cite%
{latora1} to the $q$-generalized KS entropy studied in \cite{grigolini3}.
The identity $S_{q}/t=\lambda _{q}^{(l)}$ holds for $t\rightarrow \infty $
as the interval length (around $x_{in}$) vanishes. It fluctuates, but as
explained, we look for the detailed dependence on both $x_{in}$ and $t$. The
invariant density is \textit{not} smooth and is built up by placing an
initial condition at each point $x_{in}$ on the attractor. In contrast to
the chaotic case there is not one identity but many, and the claim in \cite%
{grassberger1} that averages are needed for applications of Pesin's identity
seems ineffectual for nonmixing trajectories. Our results may not be
insignificant as these coincide (when $l=0$) with those in \cite{grigolini3}
where the $q$-KS entropy was considered. On the contrary, the entropies $%
H_{n}^{q}$ in \cite{grassberger1} from symbolic dynamics do not sense (i.e.
do not depend on) the universal $\alpha $ and/or the nonlinearity $z$.

\section{Hierarchical family of $q$-phase transitions}

With reference to the `rich zoo' of values for the index $q$ alluded in \cite%
{grassberger1}, there is a well-defined family \cite{robledo6} of these
within the attractor that deserves to be explained. This family is
determined by the discontinuities of Feigenbaum's trajectory scaling
function $\sigma $ (which measures the convergence of positions in period $%
2^{k}$ orbits as $k\rightarrow \infty $) \cite{schuster1} and are all
expressed in terms of the universal constants of unimodal maps (for general
nonlinearity $z>1$). Eqs. (\ref{sensitivity3}) and (\ref{sensitivity5}) and
the values of $q$ and $2-q$ given there are obtained from the largest
discontinuity in $\sigma $, that is itself related to the most crowded and
most sparse regions in the attractor. The other discontinuities lead to
expressions for $\xi _{t}$ similar to Eqs. (\ref{sensitivity3}) and (\ref%
{sensitivity5}). There is a corresponding family of pairs of Mori's $q$%
-phase transitions, each associated to orbits with common starting and
finishing positions at specific locations of the attractor. As before, the
special values for $q$ in $\xi _{t}$ are equal to those of the field
variable q in Mori's formalism at which the transitions occur \cite{robledo6}%
. Since the amplitudes of the discontinuities of $\sigma $ diminish rapidly,
there is a hierarchical structure and consideration only of the dominant
discontinuity provides a sound description of the dynamics. When all
discontinuities are considered it becomes apparent that the dynamics on the
critical attractor is constituted in its entirety by the infinite but
discrete set of $q$-phase transitions. We reproduce below some basic
expressions \cite{robledo6}.

The trajectory scaling function $\sigma $ is obtained as the $k\rightarrow
\infty $ limit of
\begin{equation}
\sigma _{k}(j)=\frac{d_{k+1,j}}{d_{k,j}},  \label{sigma1}
\end{equation}%
where the numbers $d_{k,j}$ are the so-called diameters that measure the
bifurcation forks that form the period-doubling cascade sequence \cite%
{schuster1}. The diameters are determined from the positions of the
'superstable' periodic orbits of lengths $2^{k}$, i.e. the $2^{k}$-cycles
that contain the point $x=0$ at $\overline{\mu }_{k}<\mu _{\infty }$ \cite%
{schuster1}. This function has finite (jump) discontinuities at all
rationals of the form $y_{j}=j/2^{j+1}$, and we denote the values of $\sigma
$ before and after them (omitting the subindex $k$) as $\sigma
(y_{j}^{-})=1/\alpha _{j}^{-}$ and $\sigma (y_{j}^{+})=1/\alpha _{j}^{+}$,
respectively. The numbers $\alpha _{j}^{\pm }$ are universal constants. For
the largest discontinuity $y_{j}=0$ one has $\alpha _{0}^{-}=\alpha ^{2}$
and $\alpha _{0}^{+}=-\alpha $.

The strategy employed \cite{robledo6} in determining $\xi _{t}$ from $\sigma
$ is to chose the initial and the final separation of the trajectories to be
the diameters $\Delta x_{in}=d_{k,j}$ and $\Delta x_{t}=d_{k,j+t}$, $%
t=2^{k}-1$, respectively. So, $\xi _{t}(x_{in})$, $x_{in}=x_{in}(j)$, is
obtained as
\begin{equation}
\xi _{t}(x_{in})=\lim_{k\rightarrow \infty }\left\vert \frac{d_{k,j+t}}{%
d_{k,j}}\right\vert .  \label{sensitivity6}
\end{equation}%
Notice that in the $k\rightarrow \infty $ limit $\Delta x_{in}\rightarrow 0$%
, $t\rightarrow \infty $ \textit{and} the $2^{k}$-supercycle becomes the
onset of chaos (the $2^{\infty }$-supercycle). Then, for each discontinuity
of $\sigma $ at $y_{j}$, $\xi _{t}(x_{in})$ can be written as \cite{robledo6}
\begin{equation}
\xi _{t,y_{j}}\simeq \left\vert \frac{\sigma
_{k}(y_{j}^{-})}{\sigma _{k}(y_{j}^{+})}\right\vert ^{k},\
t=2^{k}-1,\,k\gg1
 \label{sensitivity7}
\end{equation}%
For the inverse process, starting at $\Delta x_{in}=d_{k,j+t}=-d_{k,j-1}$
and ending at $\Delta x_{t}=d_{k,j}=-d_{k,j-1+t}$, with $t=2^{k}+1$ one
obtains \cite{robledo6}
\begin{equation}
\xi _{t,y_{j}}\simeq \left\vert \frac{\sigma
_{k}(y_{j}^{+})}{\sigma _{k}(y_{j}^{-})}\right\vert ^{k},\
t=2^{k}+1,\,k\gg1
\label{sensitivity8}
\end{equation}%
Therefore

\begin{equation}
\xi _{t,y_{j}}=\left\vert \frac{\alpha _{j}^{-}}{\alpha
_{j}^{+}}\right\vert ^{k},\; \;\;\xi _{t,y_{j}}=\left\vert
\frac{\alpha _{j}^{-}}{\alpha _{j}^{+}}\right\vert ^{-k},
\label{sensitivity9}
\end{equation}%
and, similarly to Eq. (\ref{sensitivity2}), the sensitivities in Eq. (\ref%
{sensitivity9}) can be re-expressed as $q$-exponentials with $q$-indexes%
\begin{equation}
q=1-\frac{\ln 2}{\ln \left\vert \alpha _{j}^{-}/\alpha _{j}^{+}\right\vert }%
,\;   Q=2-q=1+\frac{\ln 2}{\ln \left\vert \alpha _{j}^{-}/\alpha
_{j}^{+}\right\vert },  \label{qindexes1}
\end{equation}%
respectively. All other consequences described in the previous sections
follow in a straightforward manner.

It is interesting to note that discrete infinite families of values for the
index $q$ arise in different contexts, such as in the proposed $q$%
-generalization of the ordinary and L\'{e}vy-Gnedenko central limit theorems
\cite{tsallis11}.

As a check on the leading role of the most crowded and sparse regions of the
attractor in determining its dynamics we consider a two-scale Cantor set
approximation of the multifractal attractor of fractal dimension $d_{f}$.
Following a standard procedure \cite{beck1} for a trajectory $%
x_{0},x_{1},...,x_{2^{k+1}-1},\ k\;$large, on the attractor we cover it with
a set of intervals of lengths $l_{i}^{(k)}=\left\vert
x_{i}-x_{i+2^{k}}\right\vert $, $i=0,...,2^{k}-1$. The smallest length scale
$l_{\min }^{(k)}\sim \alpha ^{-2k}$ is observed in the neighborhood of $x=1$
while the largest $l_{\max }^{(k)}\sim \alpha ^{-k}$ occurs near $x=0$.
Trajectories starting in the vicinity of $x=1$ in the multifractal visit the
vicinity of $x=0$ at times of the form $t=2^{k}-1$. In the two-scale set the
equivalent trajectories of duration $t=2^{k}-1$ starting at positions
assigned to the scale $l_{\min }^{(k)}$ visit positions that correspond to
the scale $l_{\max }^{(k)}$ a number of times equal to $d_{f}t$. For these
times $\left\vert dx_{i}/dx_{0}\right\vert =$ $l_{\max }^{(k)}/l_{\min
}^{(k)}$ otherwise $\left\vert dx_{i}/dx_{0}\right\vert =1$. Use of this in
the time-averaged expansion rate \cite{grassberger1}%
\begin{equation}
\beta \ln t\equiv \frac{1}{t}\sum_{i=1}^{t=2^{k}-1}\ln \left\vert \frac{%
dx_{i}}{dx_{0}}\right\vert ,\ k\gg1
\label{expansionrate1}
\end{equation}%
leads to%
\begin{equation}
\beta \ln t=d_{f}\ln \frac{l_{\max }^{(k)}}{l_{\min }^{(k)}},k\gg1
\label{expansionrate2}
\end{equation}%
or
\begin{equation}
\beta =d_{f}\ln \alpha /\ln 2.  \label{beta}
\end{equation}%
As $d_{f}\simeq 0.5388$ \cite{schuster1} we obtain $\beta \simeq 0.7131$
which falls (given the number of digits considered in the value of $d_{f}$)
within $0.033\%$ of the numerical value reported for $\beta $ in \cite%
{grassberger1}. Eq. (\ref{beta}) is indeed a very simple relation linking
the expansion rate constant $\beta $ and the universal constant $\alpha $.

\section{$q$-statistics and the other routes to chaos}

With reference to the generality of our results for the Feigenbaum
attractor, and besides all unimodal maps of arbitrary nonlinearity $z>1$,
there is a very similar picture obtained for another important multifractal
critical attractor. This corresponds to the quasiperiodic route to chaos,
recently studied \cite{robledo7} in the framework of the familiar
golden-mean (gm) onset of chaos in the critical circle map \cite{schuster1},
\cite{hilborn1}. The dynamics on the attractor (a fat fractal \cite{farmer1}%
) is more involved than that for the Feigenbaum case but there is a strong
parallelism that can be glimpsed through the following equivalences when
only the most crowded and most sparse attractor regions are considered \cite%
{robledo7}:

i) The time sequences along which the sensitivity to initial conditions
exhibits the power law scaling $\xi _{t}(x_{in}=1)=\alpha ^{k}$ are $%
n=(l-m)F_{k}+mF_{k-2}$, where $F_{k}$ is the Fibonacci number of order $k$,
each sequence is obtained by running through $k=1,2,3,...$ for fixed values
of $l=1,2,3,...$ and $m=0,1,2,...,l-1$. Notice that the sequences now depend
on one additional index. As before there is a time shift $t=n-t_{w}$, here
with $t_{w}=l-m+mw_{gm}^{2}$, linking the two scales $t$ and $n$ and with
the property that rescaling of the form $t/t_{w}$ makes all power law
sequences collapse into a single one. The number $\alpha $ is now the
absolute value of the universal constant $\alpha _{gm}\simeq -1.288575$
obtained as the scaling factor that satisfies the fixed-point map equation $%
g(x)=\alpha _{gm}g(\alpha _{gm}g(x/\alpha _{gm}^{2})$.

ii) The role of the period doubling time scale factor $2^{k}$ is now taken
(asymptotically) by $w_{gm}^{-k}$ where $w_{gm}^{-1}=(\sqrt{5}-1)/2$ is the
golden mean.

iii) The sensitivity is given by $q$-exponentials as in Eqs. (\ref%
{sensitivity3}) and (\ref{sensitivity5}) with $q=1+\ln w_{gm}/2\ln \alpha $
, $\lambda _{q}^{(l,m)}=2\ln \alpha /((l-m)+mw_{gm}^{2})\ln w_{gm}$, $Q=2-q$
and $\lambda _{Q}^{(l,m)}=-2\lambda _{q}^{(l,m)}$.

iv) A pair of $q$-phase transitions take place at $\mathsf{q}=q$ and at $%
\mathsf{q}=$ $Q=2-q$, that correspond to switching starting and finishing
orbital positions. And, as before, linear growth (or reduction) of the
Tsallis entropy $S_{\mathsf{q}}$ occurs when $\mathsf{q}=q$ (or $\mathsf{q}%
=Q $) with rates $K_{q}^{(l,m)}=\lambda _{q}^{(l,m)}$ (or $%
K_{Q}^{(l,m)}=\lambda _{Q}^{(l,m)}$).

More generally, use of the trajectory scaling function $\sigma $ for the
golden mean quasiperiodic attractor reveals an infinite family of pairs of $%
q $-phase transitions, with a hierarchical structure, each member with
properties like in i-iv above \cite{robledo7}. As before the dominant
behavior arises from the most crowded and most sparse regions of the
multifractal.

Furthermore, it is significant to call attention to the fact that the
fixed-point map solution of $g(x)=\alpha g(g(x/\alpha ))$ for the tangent
bifurcations of unimodal maps of general nonlinearity $z>1$, the third route
to chaos, is rigorously given by a $q$-exponential map. See \cite{robledo1},
\cite{robledo2}. This feature leads immediately to a $q$-exponential $\xi
_{t}$ with $q=3/2$ for all $z>1$ \cite{robledo2}. The tangent bifurcations
display weak insensitivity to initial conditions, i.e. power-law convergence
of orbits when at the left-hand side ($x<x_{c}$) of the point of tangency $%
x_{c}$. However at the right-hand side ($x>x_{c}$) of the bifurcation there
is a `super-strong' sensitivity to initial conditions, i.e. a sensitivity
that grows faster than exponential \cite{robledo2}. The two different
behaviors can be couched as a $q$-phase transition with indexes $q$ and $2-q$
for the two sides of the tangency point. Also a two time or aging scaling
property like that in Eq. (\ref{sensitivity4}) holds for this critical
attractor \cite{robledo11}. The sensitivity $\xi _{t}$ is dependent on the
initial position $x_{in}$ or, equivalently, on its waiting time $t_{w}$; the
closer $x_{in}$ is to the point of tangency the longer $t_{w}$ but the
sensitivity of all trajectories fall on the same $q$-exponential curve when
plotted against $t/t_{w}$. It is worthwhile to underscore that in this case
the $q$-exponential is \textit{not} an alternative way to express a power
law but the exact function that describes the faster than exponential
increase of $\xi _{t}$.

\section{Conclusions}

Thus, we have explained in detail the dynamics at the Feigenbaum and other
critical attractors. In all cases the fluctuating sensitivity to initial
conditions has the form of infinitely many interlaced $q$-exponentials that
fold into a single one with use of the two-time scaling $t/t_{w}$ property.
More precisely, there is a hierarchy of such families of interlaced $q$%
-exponentials; an intricate (and previously unknown) state of affairs that
befits the rich scaling features of a multifractal attractor. Therefore, the
comment in the abstract of Ref. \cite{grassberger1} that "the... behavior at
the Feigenbaum point based on non-extensive thermodynamics... can be easily
deduced from well-known properties of the Feigenbaum attractor" is somewhat
an overstatement.

Summing up, the incidence of $q$-statistics in the temporal scaling at the
Feigenbaum attractor is verified; the $q$-exponential sensitivity is
explained; the origin of the index $q$ is made clear; and the equality
between $q$-generalized Lyapunov coefficients and $q$-entropy growth rates
is demonstrated. A similar corroboration applies to the other two known
routes to chaos, those via quasiperiodicity and via intermittency.\bigskip

\textbf{Acknowledgments}. Partially supported by CONACyT and DGAPA-UNAM,
Mexican agencies.

\end{document}